\documentclass[
a4paper,11pt
]{article}

\usepackage{amsfonts,amssymb}  
\usepackage{theorem}  
  
\newcommand{\der}[2]{\frac{\delta #1}{\delta #2}}

\newtheorem{theorem}{Theorem}
\newtheorem{lemma}{Lemma}

\newcommand{\lem}[1]{Lemma~{\bf\ref{lem::#1}}}    
\newcommand{\teo}[1]{Theorem~{\bf\ref{teo::#1}}}    

\newcommand\hepth[1]{{\tt hep-th/#1}}
\newcommand\plb[3]{{\em Phys.Lett.~} {\bf B} #1 (#2) #3}
\newcommand\cmp[3]{{\em Comm.Math.Phys.~} {\bf #1} (#2) #3}
\newcommand\prep[3]{{\em Phys.Rept.~} {\bf #1} (#2) #3}
\newcommand\jhep[3]{{\em Journal of High Energy Physics~} {\bf #1} (#2) #3}
\newcommand\lmp[3]{{\em Lett.Math.Phys.~} {\bf #1} (#2) #3}
\newcommand\pan[3]{{\em Phys.Atom.Nucl.~} {\bf #1} (#2) #3} 
\newcommand\prd[3]{{\em Phys.Rev.~}{\bf D #1} (#2) #3}
\newcommand\yf[3]{{\em Yad.Fiz.~}{\bf #1} (#2) #3}
\newcommand\npb[3]{{\em Nucl.Phys.~}{\bf B #1} (#2) #3}
\newcommand\ap[3]{{\em Annals Phys.~}{\bf #1} (#2) #3}

\begin{document}     

\rightline{{\bf IFUM 693/FT}}

\vskip 2 truecm

\huge
{\bf
\centerline{Algebraic Properties}
\centerline{of BRST Coupled Doublets.}
}

\normalsize
\vskip 0.7 truecm
{\Large
\centerline{Andrea Quadri \footnote{{\tt andrea.quadri@mi.infn.it}}}
}
\vskip 0.4 truecm
{\em
\centerline{Universit\`a degli Studi di Milano}
\centerline{and}
\centerline{INFN, Sezione di Milano}
\centerline{via Celoria 16, I20133 Milano - Italy}
}

\vskip 0.8 truecm
\centerline{{\bf Abstract}}
\vskip 0.7 truecm

\begin{quotation}
We characterize the dependence on doublets of 
the cohomology of an arbitrary nilpotent
differential $s$ (including BRST differentials 
and classical linearized Slavnov-Taylor (ST) operators) 
in terms of the cohomology of the doublets-independent
component of $s$.
All cohomologies are computed in the space of integrated local formal power series.
We drop the usual assumption that the counting operator
for the doublets commutes with $s$ (decoupled doublets)
and discuss the general case where the counting operator
does not commute with $s$ (coupled doublets).
The techniques used are purely algebraic and do not rely on
power-counting arguments. The main result is that the full cohomology
that includes the doublets can be obtained directly from the cohomology
of the doublets-independent component of $s$. This turns out to be
a very useful property in many problems in Algebraic Renormalization.
\end{quotation}

\newpage

\section{Introduction}

The use of BRST doublets is by far a well-established technique in
Algebraic Renormalization \cite{ps,Barnich:2000zw}. 
In gauge theories the dependence
of Green functions of gauge-invariant operators on the gauge
parameter 
\cite{Caswell:cj,Kluberg-Stern:rs,Kluberg-Stern:1974xv,Kluberg-Stern:1975hc} can be analyzed in an algebraic way by allowing the
gauge parameter to vary as a BRST doublet together with its anticommuting partner \cite{Piguet:1984js}.
Doublets are also a useful tool in proving
the independence of gauge-invariant observables of the background gauge
field \cite{Grassi:1995wr,Grassi:1999nb,Becchi:1999ir,Ferrari:2000yp}.

A couple of variables $(z,w)$ is said a doublet under the nilpotent differential
 $s$ if 
\begin{eqnarray}
s z = w \, , ~~~~  s w = 0 \, .
\label{dp::e1}
\end{eqnarray}
In the physically relevant cases $s$ is usually identified with 
the classical BRST differential or the linearized classical 
Slavnov-Taylor (ST) operator 
${\cal S}_0$ \cite{ps,Barnich:2000zw,BRST,HK}.

When the counting operator for the variables $(z,w)$ commutes with $s$
it is an easy task to show that the cohomology of $s$ is
doublets-independent. In the general case the problem is more intricated
and deserves a careful algebraic investigation.

In this paper we provide a comprehensive discussion of the dependence 
of the cohomology
of $s$ on doublets 
(also known as contractible pairs  \cite{Barnich:2000zw})
in the space of  integrated local
formal power series. We first review the standard result referring to the case of
decoupled doublets, where the counting operator for the set of doublets $(z_k,w_k)$ 
\begin{eqnarray}
{\cal N} \equiv \int d^4x \,  \sum_k \left ( z_k {\delta \over \delta z_k} + 
w_k {\delta \over \delta w_k} \right )
\label{dp::e3}
\end{eqnarray} 
commutes with the differential $s$.
In this case it can be shown
\cite{ps,Barnich:2000zw} that the contribution
of the doublets $(z_k,w_k)$ to the cohomology of $s$
is always trivial: if ${\cal I}[z,w,\varphi]$ is any $s$-invariant
integrated local formal power series in its arguments and
their derivatives, depending on the set of doublets 
$z=\{ z_k \}$, $w=\{ w_k \}$ and on a set of other fields and external sources collectively
denoted by $\varphi$,
then there exists an integrated local formal power series ${\cal G}[z,w,\varphi]$ such that
\begin{eqnarray}
{\cal I}[z,w,\varphi] - {\cal I}[0,0,\varphi] = s {\cal G}[z,w,\varphi] \, .
\label{dp::e4}
\end{eqnarray}
From the point of view of Renormalization theory, eq.(\ref{dp::e4}) implies
that the anomalies of a model, to be identified with the non-trivial
cohomology classes of the classical linearized ST operator ${\cal S}_0$
in the sector with Faddeev-Popov (FP) charge $+1$, are not affected
by the introduction of a set of decoupled doublets.

In Algebraic Renormalization decoupled doublets play for instance
an important r\^ole 
in the cohomological analysis of gauge theories, in order to prove
the independence of the cohomology of the classical linearized
ST operator ${\cal S}_0$ of antighosts and auxiliary fields
\cite{ps,Barnich:2000zw}, as well as
in the off-shell formulation of the Equivalence
Theorem \cite{ourET}.

We wish to point out that eq.(\ref{dp::e4}) 
 holds regardless the FP charge of ${\cal I}$:
it may also happen that ${\cal I}$ is a sum of 
terms
with different FP charges, although this does not happen in
Renormalization theory.

We then relax the assumption that $s$ and ${\cal N}$ commute and
discuss the more general case of coupled doublets, for which
\begin{eqnarray}
[s,{\cal N} ] \not = 0 \, .
\label{dp::e5}
\end{eqnarray}
Under suitable assumptions, in some cases \cite{brandt,Brandt:2001tg} it is known
that
one can reduce the problem
of handling coupled doublets to the decoupled case by means of a 
properly chosen change of coordinates in the jet space where $s$ acts.

It is the purpose of this paper to give a full
characterization of the dependence of the cohomology
of $s$ on coupled doublets.

In order to discuss the problem on general grounds 
it is convenient to decompose the nilpotent differential $s$
according to the degree induced by ${\cal N}$:
\begin{eqnarray}
s = \sum_{j=0}^\infty s^{(j)}
\label{E1}
\end{eqnarray}
where $s^{(j)}$ is the component of $s$ of degree $j$. Moreover we split
the zero-th order operator $s^{(0)}$ as 
\begin{eqnarray}
s^{(0)} = \bar s^{(0)} + \int d^4x \, w_k \frac{\delta}{\delta z_k} 
\label{E2}
\end{eqnarray}
where $\bar s^{(0)}$ is $(z,w)$-independent and only 
acts on the variables $\varphi$. From the nilpotency of $s$
we get that $s^{(0)}$ is nilpotent. This implies together with
eq.(\ref{E2}) that $\bar s^{(0)}$ is also nilpotent.

We will prove along the lines of homological
perturbation theory \cite{Henneaux:ig,Fisch:1989rp,Fisch:dq}
  that the cohomology
of the nilpotent differential $s$ in the space of  integrated local
formal power series in $\{ \varphi, z, w \}$ and their derivatives
is isomorphic to the cohomology of $\bar s^{(0)}$ in the space
of integrated local formal power series which depend only on
$\varphi$ and their derivatives. 

To this extent this result establishes
the independence of the cohomology of $s$ of
(generally coupled) doublets.

In practical applications
independence of the cohomology of 
doublets proves to be an important tool in the computation of cohomologies, since it allows
to restrict the cohomological problem to a smaller space
where all variables entering as doublets have been discarded.
This property has been extensively used in the literature whenever dealing with
decoupled doublets \cite{ps,Barnich:2000zw}. 
Theorem {\bf \ref{teo::new::1}}
allows to extend the range of applicability of such a technique
to the wider class of coupled doublets.

We stress that we do not rely on power-counting arguments.
This in turn allows to apply the results of the present paper 
to generally 
non power-counting renormalizable theories, as for instance in the 
BRST approach to the Equivalence Theorem discussed in 
\cite{tyutin} or in the on-shell formulation of the Equivalence Theorem \cite{ourET},
where independence of the cohomology of the linearized classical ST operator
of coupled doublets guarantees that the relevant ST identities
are anomaly-free and can thus be directly imposed order by order
in the loop expansion \cite{fg,fgq}.

\vskip 0.4 truecm

The paper is organized as follows. 
The algebraic properties
of decoupled doublets are briefly reviewed in sect.~\ref{dp::doppietti_sez2}.
The main results of the paper on coupled doublets are provided
in sect.~\ref{dp::doppietti_sez3}. Finally conclusions are reported in 
sect.~\ref{conclusions}.

\section{Decoupled doublets} \label{dp::doppietti_sez2}

In this section we review the standard result 
\cite{ps,Barnich:2000zw} showing that
the contribution of doublets to the cohomology of the differential
$s$ is trivial if
\begin{eqnarray}
[ s, {\cal N}] = 0 \, .
\label{dp::s2.1}
\end{eqnarray}
The counting operator ${\cal N}$ has been defined
in eq.(\ref{dp::e3}).

The differential $s:\Sigma \rightarrow \Sigma$ 
is assumed to act on the space $\Sigma$ of local integrated
formal 
power series depending on the set of doublets $(z,w)$, on other variables 
collectively denoted by $\varphi$ and on their derivatives.
$\Sigma$ is a linear space generated by all possible integrated linearly independent
local monomials in $\{ z,w,\varphi\}$ and their derivatives, with no restrictions on the
dimension of the generators. 

Any integrated local formal power series ${\cal G} \in \Sigma$ can be decomposed as 
\begin{eqnarray}
{\cal G} = \sum_i \int d^4x \, c_i {\cal M}_i(x) \, ,
\label{intro1}
\end{eqnarray}
where 
$\{ {\cal M}_i(x) \}$ is  a basis of local linearly independent monomials
in $\{ z,w,\varphi\}$ and their derivatives
and $c_i$ are c-number coefficients.
$\{ {\cal M}_i(x) \}$ includes monomials of arbitrarily high dimension.

The differential $s$ acting on $\Sigma$ can be expressed as
\begin{eqnarray}
s = \int d^4x \, s \varphi(x) \frac{\delta}{\delta \varphi(x)} 
   +\int d^4x \, w(x) \frac{\delta}{\delta z(x)} \, .
\label{intro2}
\end{eqnarray}

$s \varphi(x)$ is the $s$-variation of $\varphi(x)$ and is assumed to be a local
formal power series in $\{ \varphi,z,w \}$ and their derivatives, whose decomposition on
the basis $\{ {\cal M}_i(x) \}$ is given by
\begin{eqnarray}
s \varphi(x) = \sum_i d_i {\cal M}_i(x) 
\label{intro3}
\end{eqnarray}
where $d_i$ are  c-number coefficients. 

All functional derivatives are assumed to act
from the left. We also require that $s$ is nilpotent: $s^2=0$.

In practical applications $s$ is usually identified with the BRST differential 
or the classical linearized ST operator ${\cal S}_0$ \cite{ps,Barnich:2000zw}.

The space $\Sigma$ can be graded according to the counting operator in eq.(\ref{dp::e3}):
\begin{eqnarray}
\Sigma = \Sigma^{(0)} + \Sigma^{(1)} + \Sigma^{(2)} + \dots
\label{intro4}
\end{eqnarray}
where $\Sigma^{(j)}$ is the eigenspace of eigenvalue $j$ for ${\cal N}$:
\begin{eqnarray}
{\cal G} \in \Sigma^{(j)} \Rightarrow {\cal N}{\cal G} = j {\cal G} \, .
\label{intro5}
\end{eqnarray}
The action of the differential $s$ is compatible with the grading induced by ${\cal N}$
if
\begin{eqnarray}
[s, {\cal N} ] =0 \, .
\label{intro6}
\end{eqnarray}

If eq.(\ref{intro6}) is verified, we say that we are dealing with
``decoupled doublets''. This definition is motivated by the observation that
if eq.(\ref{intro6}) holds true, then no doublet $(z_k,w_k)$ can appear
in the $s$-transformation of any other field.

We follow \cite{Zumino} and  introduce the operator
\begin{eqnarray}
{\cal K} = \int_0^1 dt \, \sum_i
z_i \lambda_t {\delta \over \delta w_i} \, ,
\label{dp::k6}
\end{eqnarray}
where the operator $\lambda_t$ acts as follows
\begin{eqnarray}
\lambda_t X[z,w, \varphi] = X[t z, t w, \varphi] \, .
\label{dp::k7}
\end{eqnarray}
In the previous equation $\varphi$ denotes any set of fields
and external sources other than $(z,w)$ on which
the integrated local formal power series $X$ might depend.
By explicit computation it can be verified that ${\cal K}$
is a contracting homotopy for $s$, since it fulfills
the following equation
\begin{eqnarray}
\{ s, {\cal K} \} X = {\bf \iota} X \, .
\label{dp::k8}
\end{eqnarray}
${\bf \iota}$ is the projector on the orthogonal complement
to the kernel of ${\cal N}$:
\begin{eqnarray}
{\bf \iota} = \left . 1 \right |_{\Sigma^{(0) \, \perp}} \oplus \left . 0 \right 
|_{\Sigma^{(0)}} \, ,
\label{dp::k9}
\end{eqnarray}
where $\Sigma^{(0)}$ is
 the kernel of the counting operator ${\cal N}$ in eq.(\ref{dp::e3}) and
$\Sigma^{(0) \, \perp}$  its orthogonal complement.

Assume now that $s {\cal I} = 0$. 
${\cal I}$ can depend on $(z,w)$ and $\varphi$.
We apply eq.(\ref{dp::k8}) to ${\cal I}$ and obtain
\begin{eqnarray}
\{ s, {\cal K} \} {\cal I}[z,w,\varphi] = 
s ({\cal K} {\cal I}) = {\bf \iota} {\cal I}[z,w,\varphi]=
{\cal I}[z,w,\varphi] - {\cal I}[0,0,\varphi] \, .
\label{dp::k10}
\end{eqnarray}
In the previous equation we have used the fact that 
${\cal I}[z,w,\varphi]$ is $s$-invariant.
From eq.(\ref{dp::k10}) we conclude that the $(z,w)$-dependent
part ${\cal I}[z,w,\varphi] - {\cal I}[0,0,\varphi]$ 
of ${\cal I}$ is a $s$-exact term.

\section{Coupled doublets} \label{dp::doppietti_sez3}

Let us now move to the ``coupled'' case where
\begin{eqnarray}
[s, {\cal N}] \not= 0 \, .
\label{dp::k11}
\end{eqnarray}
In this case the operator ${\cal K}$  defined in eq.(\ref{dp::k6}) 
is no more a contracting homotopy for $s$.

We denote by $\varphi = \{ \varphi_i \}$ all the variables 
different from $(z,w)$ on which $s$ may act. We also adopt 
a compact notation and assume that
the integration over $d^4x$ is implicit in the sum over
repeated indices, i.e. we write
\begin{eqnarray}
\varphi_i {\delta X \over \delta \varphi_i} \equiv
\int d^4x \, \sum_i \varphi_i(x) {\delta X \over \delta \varphi_i(x)} \, .
\label{dp::de3}
\end{eqnarray}

Then we get the following 
decomposition for $s$:
\begin{eqnarray}
s  = g_i {\delta \over \delta \varphi_i}
+ w_i{\delta \over \delta z_i} \, .
\label{dp::de2}
\end{eqnarray}

$g_i[\varphi,z,w]$ is the $s$-variation of $\varphi_i$. 

We can decompose $s$ according to the degree induced by ${\cal N}$:
\begin{eqnarray}
s = \sum_{j=0}^\infty s^{(j)}
\label{E3}
\end{eqnarray}
where $s^{(j)}$ is the component of $s$ of degree $j$.
By comparison with eq.(\ref{dp::de2}) we have explicitly
\begin{eqnarray}
&& s^{(0)} = g_i^{(0)} \frac{\delta}{\delta \varphi_i} + w_i \frac{\delta}{\delta z_i} \nonumber \\
&& s^{(j)} = g^{(j)}_i \frac{\delta}{\delta \varphi_i} \, , 
~~~~~~~~~~~~~~~~~~~~~~~~~
j=1,2,\dots
\label{E4}
\end{eqnarray}
In the above equation $g^{(j)}_i$ denotes the component of $g_i$ in the
eigenspace of eigenvalue $j$ for the counting operator ${\cal N}$:
\begin{eqnarray}
{\cal N} g^{(j)}_i = j g^{(j)}_i \, .
\label{E5}
\end{eqnarray}
$g_i^{(0)}$ is the $(z,w)$-independent component of $g_i$.
By comparison with eq.(\ref{E2}) we see that
\begin{eqnarray}
\bar s^{(0)} = g_i^{(0)} \frac{\delta}{\delta \varphi_i} \, .
\label{E6}
\end{eqnarray}

\vskip 0.3 truecm
In most cases $s$ is to be identified with the classical linearized ST operator
${\cal S}_0$. 
We denote by $S[\phi_i,\phi_i^*,z,w]$ the classical action from which 
we define
the symplectic gradient
\begin{eqnarray}
 (S,\cdot) \equiv {\delta S \over \delta \phi_i}{\delta \over \delta \phi_i^*}
+ {\delta S \over \delta \phi^*_i} {\delta \over \delta \phi_i} \, .
\label{Symp1}
\end{eqnarray}
$S[\phi_i,\phi_i^*,z,w]$ depends on both the fields $\phi_i$
and the associated antifields $\phi_i^*$ as well as
on the doublets $(z_k,w_k)$.
The full classical linearized ST operator is then
\begin{eqnarray}
{\cal S}_0 = (S,\cdot) + w_k{\delta \over \delta z_k} = {\delta S \over \delta \phi_i}{\delta \over \delta \phi_i^*}
+ {\delta S \over \delta \phi^*_i} {\delta \over \delta \phi_i} +
 w_k {\delta \over \delta z_k} \, .
\label{dp::INTRO}
\end{eqnarray}
Nilpotency of ${\cal S}_0$
follows from the ST identity for $S$ \cite{ps}:
\begin{eqnarray}
{1 \over 2} (S,S)+ w_k {\delta S \over \delta z_k} = 0 \, .
\label{dp::beta}
\end{eqnarray}

The symplectic structure in eq.(\ref{dp::INTRO}) is not essential to prove
the results of the present section. 
In particular, we will only rely on eq.(\ref{dp::de2}), which is more
general than eq.(\ref{dp::INTRO}).

However, the results presented in this section can be derived
in an effective geometrical way \cite{STORA::private}
if the nilpotent differential $s$ 
is given by the classical linearized ST operator ${\cal S}_0$ in eq.(\ref{dp::INTRO}). 

\vskip 0.3 truecm

We will prove in Theorem {\bf \ref{teo::new::1}}
that the cohomology
of the nilpotent differential $s$ in the space of  integrated local
formal power series in $\{ \varphi, z, w \}$ and their derivatives
is isomorphic to the cohomology of $\bar s^{(0)}$ in the space
of  integrated local formal power series which depend only on
$\varphi$ and their derivatives. 

\vskip 0.3 truecm
We first show that if ${\cal I}[z,w,\varphi]$ is a 
$s$-closed  integrated local formal power series  such that 
$s{\cal I}[0,0,\varphi]=0$ then
its $(z,w)$-dependent part ${\cal I}[z,w,\varphi] -{\cal I}[0,0,\varphi]$
is $s$-exact.

\vskip 0.2 cm

\begin{lemma}\label{lem::dp::a}
Let ${\cal I}[z,w,\varphi]$ be an integrated local formal power series closed under the nilpotent
differential $s$, i.e. it fulfills the Wess-Zumino consistency
condition
\begin{eqnarray}
s {\cal I} =0 \, .
\label{dp::de3bis}
\end{eqnarray}
Moreover let us assume that $s{\cal I}[0,0,\varphi]=0$. 
Then we have 
\begin{eqnarray}
{\cal I}[z,w,\varphi] - {\cal I}[0,0,\varphi] = s {\cal G}[z,w,\varphi]
\label{dp::B1}
\end{eqnarray}
for some  integrated local formal power series ${\cal G}[z,w,\varphi]$.
\end{lemma}

\vskip 0.2 truecm
{\bf Proof.}  
The condition in eq.(\ref{dp::de3bis}) implies
\begin{eqnarray}
w_i \der{{\cal I}}{z_i} = -g_i \der{{\cal I}}{\varphi_i} \, .
\label{dp::de4}
\end{eqnarray}
Differentiating eq.(\ref{dp::de4}) with respect to $w_k$ we get
\begin{eqnarray}
\der{{\cal I}}{z_k} = (-1)^{\epsilon(w_k)+1} s \left ( \der{{\cal I}}{w_k} \right ) - 
\der{g_i}{w_k}\der{{\cal I}}{\varphi_i} \, ,
\label{dp::de5old}
\end{eqnarray}
so that eq.(\ref{dp::de5old}) becomes ($z_k$ and $w_k$ have opposite
statistics)
\begin{eqnarray}
\der{{\cal I}}{z_k} = (-1)^{\epsilon(z_k)} s \left ( \der{{\cal I}}{w_k} \right ) - 
\der{g_i}{w_k}\der{{\cal I}}{\varphi_i} \, .
\label{dp::de5}
\end{eqnarray}
In the previous equations we 
have denoted by $\epsilon(X)$ the 
Grassmann
parity of $X$ ($\epsilon(X)=0$ if
$X$ is bosonic, $\epsilon(X)=1$ if $X$ is fermionic).

We apply to both sides of eq.(\ref{dp::de5})
the operator $\int_0^1 dt \,z_k  \lambda_t$, where
the action of $\lambda_t$ on the  integrated local formal power series
 $X$ is defined by eq.(\ref{dp::k7}).

We get:
\begin{eqnarray}
\int_0^1 dt \,  z_k \lambda_t \der{{\cal I}}{z_k} = 
\int_0^1 dt \, \left ( (-1)^{\epsilon(z_k)} z_k \lambda_t 
s \left ( \der{{\cal I}}{w_k} \right ) - 
z_k \lambda_t \der{g_i}{w_k}\der{{\cal I}}{\varphi_i} \right ) \, .
\label{dp::de7}
\end{eqnarray}
On the other hand 
\begin{eqnarray}
\int_0^1 dt \,  (-1)^{\epsilon(z_k)} z_k \lambda_t 
s \left ( \der{{\cal I}}{w_k} \right ) & = & 
\int_0^1 dt \, \left (  (-1)^{\epsilon(z_k)} 
z_k s \left ( \lambda_t \der{{\cal I}}{w_k} \right ) \right . \nonumber \\
&&  ~~~~~~~~ \left .
- (-1)^{\epsilon(z_k)} z_k [s, \lambda_t] \der{{\cal I}}{w_k} \right ) \nonumber \\
& = & \int_0^1 dt \, 
\left( s \left (z_k \lambda_t  \der{{\cal I}}{w_k} \right ) - 
w_k \lambda_t \der{{\cal I}}{w_k} \right . \nonumber \\
& & ~~~~~~~~
\left . - (-1)^{\epsilon(z_k)} z_k [s, \lambda_t] \der{{\cal I}}{w_k} \right )
\, .
\label{dp::de8}
\end{eqnarray}
Substituting in eq.(\ref{dp::de7}) yields
\begin{eqnarray}
&& \!\!\!\!\!\!\!\!\!\!\!\!\!
\!\!\!\!\! \int_0^1 dt \, \left ( z_k \lambda_t \der{{\cal I}}{z_k} +
w_k \lambda_t \der{{\cal I}}{w_k} \right ) = \nonumber \\
&& \!\!\!\!\!\!
\int_0^1 dt \, \left (  s \left (z_k \lambda_t  \der{{\cal I}}{w_k} \right )
-z_k \left ( (-1)^{\epsilon(z_k)} [s, \lambda_t] \der{{\cal I}}{w_k}
+\lambda_t \der{g_i}{w_k}\der{{\cal I}}{\varphi_i} \right ) 
\right ) \, .
\label{dp::de9}
\end{eqnarray}
The L.H.S. in eq.(\ref{dp::de9}) gives
${\cal I}[z,w,\varphi] - {\cal I}[0,0,\varphi]$, so that
\begin{eqnarray}
\!\!\!\!\!\!\!\! {\cal I}[z,w,\varphi] - {\cal I}[0,0,\varphi] & = &
\int_0^1 dt \, \left [ s \left (z_k \lambda_t  \der{{\cal I}}{w_k} \right )
\right . \nonumber \\
&  &  \left . -z_k \left (  (-1)^{\epsilon(z_k)} [s, \lambda_t] \der{{\cal I}}{w_k}
+\lambda_t \der{g_i}{w_k}\der{{\cal I}}{\varphi_i} \right ) 
\right ] \, .
\label{dp::de10}
\end{eqnarray}
Notice that 
\begin{eqnarray}
[s ,\lambda_t] = [g_i \der{(\cdot)}{\varphi_i},\lambda_t] \, .
\label{dp::de11}
\end{eqnarray}
Let us define now 
\begin{eqnarray}
{\cal I}_1 = - \int_0^1 dt \, z_k \left (  (-1)^{\epsilon(z_k)}
[s, \lambda_t] \der{{\cal I}}{w_k}
+\lambda_t \der{g_i}{w_k}\der{{\cal I}}{\varphi_i} \right ) \, .
\label{dp::de12}
\end{eqnarray}
By using $s {\cal I} =0$, 
$s {\cal I}[0,0,\varphi]=0$ and 
the fact that $s^2=0$ 
we get from  eq.(\ref{dp::de10})
\begin{eqnarray}
s {\cal I}_1 =0 \, .
\label{dp::de13}
\end{eqnarray}
Moreover we see from eq.(\ref{dp::de12}) that ${\cal I}_1$ satisfies the condition
$$\left . {\cal I}_1 [z,w,\varphi] \right |_{z=w=0} = 0.$$
Thus in particular
\begin{eqnarray}
s {\cal I}_1[0,0,\varphi] = 0 \, .
\label{dp::B2}
\end{eqnarray}
Hence we can repeat the argument used for ${\cal I}$ and write 
${\cal I}_1$ in the form
\begin{eqnarray}
&&\!\!\!\!\!\!\!\!\!\!\!\!\!\!\!\!\!\!  {\cal I}_1[z,w,\varphi] =  
{\cal I}_1[z,w,\varphi] - {\cal I}_1[0,0,\varphi] \nonumber \\
&& \!\!\!\!\!\!\!\!\!\!\!\!\! = 
\int_0^1 dt \, \left [ s \left (z_k \lambda_t  \der{{\cal I}_1}{w_k} \right )
-z_k \left ( (-1)^{\epsilon(z_k)} [s, \lambda_t] \der{{\cal I}_1}{w_k}
+\lambda_t \der{g_i}{w_k}\der{{\cal I}_1}{\varphi_i} \right ) \right ] \, .
\label{dp::de14}
\end{eqnarray}

By taking into account eq.(\ref{dp::de11}) we see that the second term in
the R.H.S. of eq.(\ref{dp::de14}) contains at least two $z$'s.

The construction can be iterated. Assume that 
${\cal I}_n$, $n \geq 1$  contains the product of at least $n$ $z$'s and assume that
$s {\cal I}_n=0$.
Moreover, we assume that $\left . {\cal I}_n[z,w,\varphi] \right |_{z=w=0} = 0$,
which implies in particular $s {\cal I}_n[0,0,\varphi]=0$.
Then we can write
\begin{eqnarray}
{\cal I}_n[z,w,\varphi] & = & {\cal I}_n[z,w,\varphi] - {\cal I}_n[0,0,\varphi] 
\nonumber \\
& = &
s \left [ \int_0^1 dt \, \left (z_k \lambda_t  \der{{\cal I}_n}{w_k}
\right )
\right ] + {\cal I}_{n+1} \, .
\end{eqnarray}
In the previous equation we have defined
\begin{eqnarray}
{\cal I}_{n+1} =  - \int_0^1 dt \, z_k \left ( (-1)^{\epsilon(z_k)} [s, \lambda_t] \der{{\cal I}_n}{w_k}
+\lambda_t \der{g_i}{w_k}\der{{\cal I}_n}{\varphi_i} \right ) \, .
\end{eqnarray}
We now notice that ${\cal I}_{n+1}$ contains the product of at least 
$(n+1)$ $z$'s.  Moreover $s {\cal I}_{n+1}=0$ 
and
$\left . {\cal I}_{n+1}(z,w,\varphi) \right |_{z=w=0}=0$.

Thus we obtain that ${\cal I}$ is $s$-exact up to a term
containing $(n+1)$ $z$'s.

By the previous arguments we can therefore conclude that 
${\cal I}[z,w,\varphi]-{\cal I}[0,0,\varphi]$
is $s$-exact as a  integrated local formal power series in $z$.
This ends the proof.

\vskip 0.3 truecm

The above argument provides an explicit although recursive 
representation of the local integrated formal power series ${\cal G}$
in eq.(\ref{dp::B1}):
\begin{eqnarray}
{\cal G} = \int_0^1 dt \, s \Big [ 
z_k \lambda_t \frac{\delta}{\delta w_k} \Bigg ( \sum_{j=0}^\infty
{\cal I}_j \Bigg ) 
\Big ] \, ,
\label{add1}
\end{eqnarray}
where we have set ${\cal I}_0 \equiv {\cal I}$.

Of course ${\cal G}$ is not unique, since for any $s$-invariant
local formal power series ${\cal F}$ we have that
\begin{eqnarray}
{\cal G}' \equiv {\cal G} + {\cal F} 
\label{add2}
\end{eqnarray}
also satisfies eq.(\ref{dp::B1}) if ${\cal G}$ does.

\vskip 0.3 truecm

In particular, if ${\cal I}[0,0,\varphi]=0$ the previous Lemma implies
that the whole $s$-invariant ${\cal I}[z,w,\varphi]$ is $s$-exact: any 
$s$-closed local integrated formal power series vanishing at $z=w=0$ is $s$-exact.

We remark that the condition
\begin{eqnarray}
s {\cal I}[0,0,\varphi] =0 
\label{dp::A1_1}
\end{eqnarray}
is also necessary if the $(z,w)$-dependence of the $s$-invariant 
${\cal I}[z,w,\varphi]$ has to be cohomologically trivial. 
This is true independently of the coupled or decoupled nature
of the doublets under investigation.
Indeed assume that there exists an integrated local formal power series
${\cal G}[z,w,\varphi]$ such that 
\begin{eqnarray}
{\cal I}[z,w,\varphi] - {\cal I}[0,0,\varphi] = s {\cal G}[z,w,\varphi] \, .
\label{dp::A2_1}
\end{eqnarray}
Then by the nilpotency of $s$ and the $s$-invariance of ${\cal I}[z,w,\varphi]$
we conclude that
\begin{eqnarray}
s {\cal I}[0,0,\varphi] = 0 \, .
\label{dp::A3_1}
\end{eqnarray}
Therefore eq.(\ref{dp::A3_1}) is both a necessary and sufficient (by virtue
of \lem{dp::a}) condition in order that the dependence of the $s$-invariant
${\cal I}[z,w,\varphi]$ on the doublets $(z,w)$ is cohomologically trivial
in the sense of eq.(\ref{dp::e4}).

Notice that in the case of decoupled doublets 
eq.(\ref{dp::A3_1}) is actually a consequence
of the fact that ${\cal I}[z,w,\varphi]$ is $s$-invariant, i.e.
of the equation
\begin{eqnarray}
s {\cal I}[z,w,\varphi]= 0 
\label{dp::A4_1}
\end{eqnarray}
once one takes the zero-th order (with respect to the
grading induced by ${\cal N}$) component of eq.(\ref{dp::A4_1}).

\lem{dp::a} turns out to be a useful tool in many practical
applications, like for instance in the discussion of the on-shell
case of the Equivalence Theorem \cite{ourET}. 

By using \lem{dp::a} we will now prove in Theorem {\bf \ref{teo::new::1}} 
the main result that
the cohomology of $s$ is independent of the (generally coupled) 
doublets $(z,w)$.
This in turn allows to characterize 
the full cohomology of  $s$ in terms of 
the cohomology of the doublets-independent component
of $s$.

\begin{theorem}
\label{teo::new::1}
The cohomology of the nilpotent differential
$s$ in the space $\Sigma$ of  integrated local formal power series
in $\{ \varphi, z, w\}$ and their derivatives is isomorphic to the
cohomology of $\bar s^{(0)}$ in the space $\Sigma^{(0)}$ of 
integrated local formal power series which only depend on $\varphi$ and
their derivatives:
\begin{eqnarray}
H(s,\Sigma) \approx H(\bar s^{(0)},\Sigma^{(0)}) \, .
\label{Eth}
\end{eqnarray}
\end{theorem}
\vskip 0.1 truecm
{\bf Proof.}
We will explicitly construct an isomorphism $\Phi$
between $H(s,\Sigma)$ and $H(\bar s^{(0)},\Sigma^{(0)})$.
Let ${\cal I} \in \Sigma$  be such that
\begin{eqnarray}
s {\cal I} = 0 \, .
\label{et1}
\end{eqnarray}
We decompose ${\cal I}$ according to the degree induced by ${\cal N}$,
i.e.
\begin{eqnarray}
{\cal I} = \sum_{k=0}^\infty {\cal I}^{(k)} 
\label{et2}
\end{eqnarray}
where ${\cal I}^{(k)}$ belongs to the eigenspace of eigenvalue $k$
of the counting operator ${\cal N}$:
\begin{eqnarray}
{\cal N} {\cal I}^{(k)} = k {\cal I}^{(k)} \, .
\label{et3}
\end{eqnarray}
We notice that
\begin{eqnarray}
\bar s^{(0)} {\cal I}^{(0)} = 0 
\label{et3bis}
\end{eqnarray}
i.e. ${\cal I}^{(0)}$ belongs to the cohomology of $\bar s^{(0)}$ in 
$\Sigma^{(0)}$. 
This follows from eq.(\ref{et1}) once we consider its zero-th order
in the expansion based on the grading induced by ${\cal N}$:
\begin{eqnarray}
(s^{(0)} + s^{(1)} + \dots) ({\cal I}^{(0)} + {\cal I}^{(1)} + \dots ) = 0
\Rightarrow s^{(0)} {\cal I}^{(0)} = 0 \, .
\label{et3ter}
\end{eqnarray}
The dots stand for term of higher order with respect to the grading
induced by ${\cal N}$.
Since ${\cal I}^{(0)}$ is independent of $(z,w)$ 
the R.H.S. of eq.(\ref{et3ter})
entails eq.(\ref{et3bis}).

Let us now set
\begin{eqnarray}
\Phi [{\cal I} ] = [{\cal I}^{(0)}]
\label{et4}
\end{eqnarray}
where $[{\cal I}]$ stands for the cohomology class of ${\cal I}$
in $H(s,\Sigma)$ and $[{\cal I}^{(0)}]$ stands for the cohomology class
of ${\cal I}^{(0)}$ in $H({\bar s}^{(0)}, \Sigma^{(0)})$.

The map in eq.(\ref{et4}) is well-defined in cohomology since if
\begin{eqnarray}
{\cal I} = s {\cal G} 
\label{et5}
\end{eqnarray}
then also
\begin{eqnarray}
{\cal I}^{(0)} = \bar s^{(0)} {\cal G}^{(0)} \, .
\label{et6}
\end{eqnarray}
This follows by expanding eq.(\ref{et5}) according to the grading
induced by ${\cal N}$ once we look at the zero-th order terms:
\begin{eqnarray}
&& {\cal I}^{(0)} + {\cal I}^{(1)} + \dots =
( s^{(0)} + s^{(1)} + \dots) ({\cal G}^{(0)} + {\cal G}^{(1)} + \dots) 
\nonumber \\
&& ~~~~~~~~~~~~~~~~~~
\Rightarrow {\cal I}^{(0)} = s^{(0)} {\cal G}^{(0)} \, .
\label{et7}
\end{eqnarray}
We recover eq.(\ref{et6}) from the above equation since ${\cal G}^{(0)}$ 
is independent of $(z,w)$.

We now show that $\Phi$ is an isomorphism.

Let us first prove that $\Phi$ is surjective. Given any  
$\bar s^{(0)}$-invariant 
${\cal I}^{(0)}$ we show that it can be completed
to a $s$-invariant ${\cal I}$. So we look for the coefficients 
${\cal I}^{(1)},{\cal I}^{(2)},\dots$ in such a way that
\begin{eqnarray}
{\cal I} \equiv {\cal I}^{(0)} + {\cal I}^{(1)} + {\cal I}^{(2)} + \dots
\label{et8}
\end{eqnarray}
fulfills 
\begin{eqnarray}
s {\cal I} = ( s^{(0)} + s^{(1)} + s^{(2)} \dots ) 
             ( {\cal I}^{(0)} + {\cal I}^{(1)} + {\cal I}^{(2)} + \dots) = 0
\, .
\label{et9}
\end{eqnarray}
The proof is a recursive one. At order zero eq.(\ref{et9}) is true since
${\cal I}^{(0)}$ is a $\bar s^{(0)}$-invariant and ${\cal I}^{(0)}$ is 
$(z,w)$-independent, so that
\begin{eqnarray}
0 = {\bar s}^{(0)} {\cal I}^{(0)} = s^{(0)} {\cal I}^{(0)} \, .
\label{et10}
\end{eqnarray}
We next move to the first order. From the nilpotency of $s$ we get
\begin{eqnarray}
s^{(1)} s^{(0)} + s^{(0)} s^{(1)} = 0 \, .
\label{et11}
\end{eqnarray}
Since $s^{(0)} {\cal I}^{(0)} = 0$ we get from the above equation
\begin{eqnarray}
s^{(0)} s^{(1)} {\cal I}^{(0)} = 0 \, .
\label{et12}
\end{eqnarray}
$s^{(0)}$ is a nilpotent differential with respect to which $(z,w)$ 
form a set of decoupled doublets. By using the standard results on the
independence of the cohomology of decoupled doublets reported in
Sect.~\ref{dp::doppietti_sez2} we conclude from eq.(\ref{et12}) that there
must exist an integrated local formal power series ${\cal I}^{(1)}$ such that
\begin{eqnarray}
s^{(1)} {\cal I}^{(0)} = s^{(0)} ( - {\cal I}^{(1)} ) \, .
\label{et13}
\end{eqnarray}
This in turn implies that eq.(\ref{et9}) is fulfilled at order one
since from eq.(\ref{et13}) we have
\begin{eqnarray}
s^{(1)} {\cal I}^{(0)} + s^{(0)} {\cal I}^{(1)} = 0 \, .
\label{et13bis}
\end{eqnarray}
The construction can be iterated to all orders. Assume that ${\cal I}$ 
has been constructed up to order $n-1$ by assigning the coefficients 
${\cal I}^{(0)}, {\cal I}^{(1)}, \dots, {\cal I}^{(n-1)}$, 
fulfilling
\begin{eqnarray}
\sum_{j=0}^m s^{(j)} {\cal I}^{(m-j)} = 0 \, , ~~~~~~~
m=0,1,\dots,n-1 \, .
\label{et14}
\end{eqnarray}
Then by eq.(\ref{et14}) we see that 
$$ s \sum_{k=0}^{n-1} {\cal I}^{(k)} $$
starts with a coefficient of order $n$, let us call it $\Delta^{(n)}$:
\begin{eqnarray}
s \sum_{k=0}^{n-1} {\cal I}^{(k)} = \Delta^{(n)} + \dots
\label{et15}
\end{eqnarray}
One has explicitly
\begin{eqnarray}
\Delta^{(n)} = \sum_{j=1}^n s^{(j)} {\cal I}^{(n-j)} \, .
\label{et15bis}
\end{eqnarray}
Since $s$ is nilpotent we get from eq.(\ref{et15})
\begin{eqnarray}
s^2 \sum_{k=0}^{n-1} {\cal I}^{(k)} = s (\Delta^{(n)} + \dots ) = 0 \, ,
\label{et16}
\end{eqnarray}
so that
\begin{eqnarray}
(s^{(0)} + s^{(1)} + \dots) (\Delta^{(n)} + \dots ) = 0 \, .
\label{et16_new}
\end{eqnarray}
We  look at the lowest order contribution to the above equation,
i.e. at order $n$, and get
\begin{eqnarray}
s^{(0)} \Delta^{(n)} = 0 \, .
\label{et17}
\end{eqnarray}
By using again the results on decoupled doublets of Sect.~\ref{dp::doppietti_sez2}
we can show that there exists an integrated local formal power series
${\cal I}^{(n)}$ such that
\begin{eqnarray}
\Delta^{(n)} = - s^{(0)} {\cal I}^{(n)} \, .
\label{et18}
\end{eqnarray}
By taking into account eq.(\ref{et15bis}) we get from eq.(\ref{et18})
\begin{eqnarray}
s^{(0)} {\cal I}^{(n)} + \sum_{j=1}^n s^{(j)} {\cal I}^{(n-j)} =
\sum_{j=0}^n s^{(j)} {\cal I}^{(n-j)} = 0 \, ,
\label{et19}
\end{eqnarray}
so that eq.(\ref{et9}) is fulfilled at order $n$.

We now show that $\Phi$ is also one-to-one by proving that
\begin{eqnarray}
{\rm ker} \, \Phi = \{ [0] \} \, .
\label{et20}
\end{eqnarray}
For that purpose let us take a $s$-invariant ${\cal I}$ such that
\begin{eqnarray}
\Phi([ {\cal I} ]) = [0] \, .
\label{et20_new}
\end{eqnarray}
This means that the
zero-th order component ${\cal I}^{(0)}$ of ${\cal I}$ fulfills
\begin{eqnarray}
{\cal I}^{(0)} = {\bar s}^{(0)} {\cal G}^{(0)}
\label{et21}
\end{eqnarray}
for some integrated local formal power series ${\cal G}^{(0)}$ independent of $(z,w)$.
Then
{\small
\begin{eqnarray}
{\cal I} - s {\cal G}^{(0)} & = & 
\left ( {\cal I}^{(0)} - {\bar s}^{(0)} {\cal G}^{(0)} \right ) +
\left ( {\cal I}^{(1)} - s^{(1)} {\cal G}^{(0)}  
\right ) 
+ \left ( {\cal I}^{(2)} - s^{(2)} {\cal G}^{(0)}  
\right ) 
+ \dots \nonumber \\
& = & \left ( {\cal I}^{(1)} - s^{(1)} {\cal G}^{(0)}  
\right ) + \left ( {\cal I}^{(2)} - s^{(2)} {\cal G}^{(0)}  
\right ) + \dots
\label{et22}
\end{eqnarray}
}
In the second line of the above equation we have used eq.(\ref{et21}).

From eq.(\ref{et22}) we see that 
$\left . \left ( {\cal I} - s {\cal G}^{(0)} \right ) \right |_{z=w=0} = 0\, .$
Moreover 
$$ {\cal I} - s {\cal G}^{(0)} $$
is $s$-invariant since $s {\cal I}=0$ and $s^2=0$.
Thus ${\cal I} - s {\cal G}^{(0)}$ fulfills the assumptions
of \lem{dp::a} and we conclude that
\begin{eqnarray}
{\cal I} - s {\cal G}^{(0)} = s {\cal H} 
\label{et23}
\end{eqnarray}
for some integrated local formal power series ${\cal H}$. Hence we see that
\begin{eqnarray}
{\cal I} = s ({\cal G}^{(0)} + {\cal H}) \, .
\label{et24}
\end{eqnarray}
From eq.(\ref{et24}) we conclude that ${\cal I}$ is $s$-exact.
Therefore  eq.(\ref{et20}) is verified and the Theorem is proven.

\vskip 0.3 truecm

We notice that
the injectivity of $\Phi$ can also be derived in an alternative fashion by using
the fact that the cohomology of $s_0$ is concentrated in degree zero.
The proof is a recursive one.  By eq.(\ref{et21}) we get
\begin{eqnarray}
{\cal I} & = & {\cal I}^{(0)} + {\cal I} - {\cal I}^{(0)} = \bar s^{(0)} {\cal G}^{(0)}
              +{\cal I} - {\cal I}^{(0)} \nonumber \\
	     & = & s {\cal G}^{(0)} - (s - \bar s^{(0)}) {\cal G}^{(0)} + {\cal I} - {\cal I}^{(0)}
		     \nonumber \\
	     & = & s {\cal G}^{(0)} + {\cal A}_1 
\label{new1}
\end{eqnarray}
where ${\cal A}_1 \equiv - (s - \bar s^{(0)}) {\cal G}^{(0)} + {\cal I} - {\cal I}^{(0)}$
is at least of order $1$ in the degree induced by ${\cal N}$.
Therefore ${\cal I}$ is $s$-exact up to order $1$.

Let us now assume that ${\cal I}$ has been shown to be $s$-exact up to order 
$k$:
\begin{eqnarray}
{\cal I} = s {\cal F}_{j-1}	+ {\cal A}_j \, , ~~~~~~~ j=1,2\dots,k
\label{new2}
\end{eqnarray}
where ${\cal A}_j$ is at least of order $j$ and can therefore be decomposed as
\begin{eqnarray}
{\cal A}_j = {\cal A}^{(j)}_j + {\cal A}^{(j+1)}_j + \dots
\label{new3}
\end{eqnarray}
In the above equation
 ${\cal A}_j^{(m)}$ is the component of order $m$ of ${\cal A}_j$.

For $j=1$ we get  ${\cal F}_0 = {\cal G}^{(0)}$, by comparison
of eq.(\ref{new1}) and  eq.(\ref{new2}).

We now prove that eq.(\ref{new2}) is also verified at order $k+1$
if it is fulfilled up to order $k$.
For that purpose we compute the $s$-variation of both sides of eq.(\ref{new2}). Since
$s^2=0$ we get at order $k$:
\begin{eqnarray}
s {\cal I} = s^2 {\cal F}_{k-1} + s {\cal A}_k = s {\cal A}_k \, .
\label{new4}
\end{eqnarray}
Since $s {\cal I}=0$ the above equation gives
\begin{eqnarray}
0 =  s {\cal A}_k \, .
\label{new5}
\end{eqnarray}
Let us expand eq.(\ref{new5}) according to the degree induced by ${\cal N}$.
The first non-zero term is of order $k$ and reads
\begin{eqnarray}
0 = s_0 {\cal A}_k^{(k)} \, .
\label{new6}
\end{eqnarray}
Since the cohomology of $s_0$ is concentrated in degree zero there exists
an integrated local formal power series ${\cal B}_k^{(k)}$ such that
\begin{eqnarray}
{\cal A}_k^{(k)} = s_0 {\cal B}_k^{(k)} \, .
\label{new7}
\end{eqnarray}
We insert eq.(\ref{new7}) into eq.(\ref{new2}) with $j=k$ and get
\begin{eqnarray}
{\cal I} & = & s {\cal F}_{k-1} + {\cal A}^{(k)}_k + {\cal A}^{(k+1)}_k 
               + {\cal A}^{(k+2)}_k + \dots \nonumber \\
	     & = & s {\cal F}_{k-1} + s_0 {\cal B}_k^{(k)} + {\cal A}^{(k+1)}_k 
               + {\cal A}^{(k+2)}_k + \dots \nonumber \\
		 & = & s ( {\cal F}_{k-1} +  {\cal B}_k^{(k)} ) -
		       (s - s_0)  {\cal B}_k^{(k)} +  {\cal A}_k  - {\cal A}_k^{(k)}
              \, . 
\label{new8}		 		        
\end{eqnarray}
The integrated local formal power series ${\cal A}_{k+1}$ given by
\begin{eqnarray}
{\cal A}_{k+1} \equiv  -
		       (s - s_0)  {\cal B}_k^{(k)} +   {\cal A}_k  - {\cal A}_k^{(k)}
\label{new9}
\end{eqnarray}
is at least of order $k+1$. If we choose
\begin{eqnarray}
{\cal F}_k \equiv {\cal F}_{k-1} + {\cal B}^{(k)}_k
\label{new10}
\end{eqnarray}
eq.(\ref{new8}) can be rewritten as
\begin{eqnarray}
{\cal I} = s {\cal F}_k + {\cal A}_{k+1} \, .
\label{new11}
\end{eqnarray}
Thus eq.(\ref{new2}) is verified at order $k+1$.
This concludes the recursive proof of the injectivity of $\Phi$.

\section{Conclusions}\label{conclusions}

In the present paper we have discussed on general grounds 
the dependence of nilpotent
differentials on doublets, both in the decoupled and the
coupled case. 

We have explicitly constructed an isomorphism between 
the cohomology of $s$ in the space of integrated local
formal power series depending on the set of doublets $(z,w)$ 
and on $\varphi$ and their derivatives and the cohomology
of $\bar s^{(0)}$ in the space of  integrated local
formal power series only depending on $\varphi$ and their
derivatives.

To this extent the cohomology of any nilpotent 
differential $s$ in the space of  integrated local formal power series
is independent of doublets both in the decoupled and the coupled case.

As a final point we remark that the whole analysis was purely
algebraic. No use was made of  power-counting
arguments.

\vskip 0.3 truecm

\section*{Acknowledgements}

Useful discussions with Prof.~Ruggero Ferrari and Prof.~Raymond Stora 
are gratefully acknowledged. The author wishes to thank 
Prof.~Marc Henneaux for many comments and suggestions and 
for his valuable help 
in the proof of \teo{new::1}.
Partial financial support from MIUR (Ministero dell'Istruzione,
dell'Universit\`a e della Ricerca Scientifica e Tecnologica) is gratefully acknowledged.

\end{document}